\documentclass{emulateapj} 
\usepackage{graphicx}
\usepackage{apjfonts}

\newcommand{\be}{\begin{equation}}
\newcommand{\ee}{\end{equation}}
\newcommand{\ba}{\begin{eqnarray}}
\newcommand{\ea}{\end{eqnarray}}

\newcommand{\mic}{\mu{\rm m}}

\newcommand{\uJy}{\mu{\rm Jy}}
\newcommand{\Msol}{\rm M_\odot}

\newcommand{\Msolpyr}{\rm M_\odot\,{\rm yr}^{-1}}
\newcommand{\kms}{\rm km\,s^{-1}}

\newcommand{\Mpc}{\rm Mpc}

\newcommand{\dex}{\rm dex}
\newcommand{\sfr}{\rm SFR}

\newcommand{\ergs}{\rm erg\,s^{-1}}
\newcommand{\fsf}{f_{SF}}
\newcommand{\Lirstar}{L_{IR}^\star}
\newcommand{\Kstar}{K^\star}

\def\ls{\mathrel{\hbox{\rlap{\hbox{\lower4pt\hbox{$\sim$}}}\hbox{$<$}}}}
\def\gs{\mathrel{\hbox{\rlap{\hbox{\lower4pt\hbox{$\sim$}}}\hbox{$>$}}}}
\shorttitle{The Mid-IR Butcher-Oemler Effect}
\shortauthors{Haines et al.}
\begin{document}
\title{LoCuSS: The Mid-Infrared Butcher-Oemler effect}

\author{
  C.\ P.\ Haines,$\!$\altaffilmark{1}
  G.\ P.\ Smith,$\!$\altaffilmark{1}
  E.\ Egami,$\!$\altaffilmark{2}
  R.\ S.\ Ellis,$\!$\altaffilmark{3,4}
  S.\ M.\ Moran,$\!$\altaffilmark{5}
  A.\ J.\ R.\ Sanderson,$\!$\altaffilmark{1}\\
  P.\ Merluzzi,$\!$\altaffilmark{6}
  G.\ Busarello,$\!$\altaffilmark{6}
  R.\ J. Smith$\!$\altaffilmark{7}
}

\altaffiltext{1}{School of Physics and Astronomy, University of
  Birmingham, Edgbaston, Birmingham, B15 2TT, UK;
  cph@star.sr.bham.ac.uk}
\altaffiltext{2}{Steward Observatory, University of Arizona, 933 North
  Cherry Avenue, Tucson, AZ 85721, USA} 
\altaffiltext{3}{California Institute of Technology, 105-24 Astronomy,
  Pasadena, CA 91125, USA} 
\altaffiltext{4}{Department of Astrophysics, University of Oxford,
  Keble Road, Oxford, OX1 3RH, UK}
\altaffiltext{5}{Department of Physics and Astronomy, The Johns
  Hopkins University, 3400 N. Charles Street, Baltimore, MD 21218,
  USA} 
\altaffiltext{6}{INAF - Osservatorio Astronomico di Capodimonte, via
  Moiariello 16, I-80131 Napoli, Italy}
\altaffiltext{7}{Department of Physics, Durham University, Durham, DH1
  3LE, UK}


\begin{abstract}
  We study the mid-infrared properties of galaxies in 30 massive
  galaxy clusters at $0.02{\le}z{\le}0.40$, using
  panoramic {\em Spitzer}/MIPS $24\mic$ and near-infrared data, including 27 new observations from the LoCuSS and ACCESS surveys. 
 This is the largest sample of
  clusters to date with such high-quality and uniform mid-infrared
  data covering not only the cluster cores, but extending into the
  infall regions. We use these data to revisit the so-called
  Butcher-Oemler effect, measuring the fraction of massive infrared
  luminous galaxies (\mbox{$K{<}\Kstar{+}1.5$},
  $L_{IR}{>}5{\times}10^{10}L_\odot$) within $r_{200}$, finding a
  steady increase in the fraction with redshift from ${\sim}3\%$ at
  $z{=}0.02$ to ${\sim}10\%$ by $z{=}0.30$, and an rms cluster-to-cluster scatter about this trend of 0.03.  
  The best-fit redshift evolution model of the form
  $\fsf\propto(1+z)^n$ has $n=5.7^{+2.1}_{-1.8}$, which is stronger
  redshift evolution than that of $\Lirstar$ in both clusters and the
  field.  We find that, statistically, this excess is associated with
  galaxies found at large cluster-centric radii, specifically
  $r_{500}{<}r{<}r_{200}$, implying that the mid-infrared Butcher Oemler
  effect can be explained by a combination of both the global decline
  in star-formation in the universe since $z\sim1$ and enhanced star
  formation in the infall regions of clusters at intermediate
  redshifts.  This picture is supported by a simple infall model based
  on the Millennium Simulation semi-analytic galaxy catalogs, whereby
  star-formation in infalling galaxies is instantaneously quenched
  upon their first passage through the cluster, in that the observed
  radial trends of $\fsf$ trace those inferred from the simulations.
  The observed $\fsf$ however lie systematically above the
    predictions, suggesting an overall excess of star-formation,
    either due to triggering by environmental processes, or a gradual
    quenching. We also find that $\fsf$ does not depend on simple
    indicators of the dynamical state of clusters, including the
    offset between the brightest cluster galaxy and the peak of the
    X-ray emission.  This is consistent with the picture described
    above in that most new star-formation in clusters occurs in the
    infall regions, and is thus not sensitive to the details of
    cluster-cluster mergers in the core regions.
\end{abstract}


\section{Introduction}
\label{intro}

\setcounter{footnote}{7}

The evolution of galaxies in clusters since $z\simeq1$ is expected to
reflect both changes in the raw ingredients, i.e.\ the properties of galaxies
that have fallen into clusters from the field in this time, and the
physical processes that have acted on those galaxies after infall.
Early evidence for cluster galaxy evolution was presented by
\citet[][hereafter BO84]{bo78,bo84}, who found that the fraction of
cluster members bluer than the cluster red sequence ($f_{b}$), by at
least $\Delta_{B-V}{=}0.2$\,mag in the $B{-}V$ rest frame, increases
from zero in the local universe to ${\sim}0.2$ by $z{\simeq}0.4$
implying a rapid evolution of the cluster population over the last 5
billion years.  Empirically the star-forming spiral galaxies found by
BO84 at $z{\sim}0.2-0.4$ are mostly replaced by S0 galaxies in local
clusters \citep{dressler97,treu}.  A simple interpretation is that
clusters accreted blue gas-rich star-forming spirals at $z{\ge}0.5-1$
and that these galaxies have been transformed somehow into the passive
S0s found in local clusters, by having their gas reservoirs depleted
by one or more physical processes within clusters, including for
example ram-pressure stripping, starvation or harassment \citep[for
reviews see][]{boselli,haines07}.

However, more recent studies of the so called Butcher-Oemler (BO)
effect and the evolution of the morphology-density relation have
suggested that at least some of the observed evolution in $f_b$ is due
to selection biases.  Firstly, BO84 selected galaxies on optical
luminosity ($M_V$) rather than stellar mass (or $M_K$), and were
therefore susceptible to biases arising from low-mass spiral galaxies
-- the optical luminosities of such galaxies are boosted by starburst
activity, and thus they increasingly enter the samples at higher
redshifts \citep{depropris03,holden}.  Similarly, the use of a fixed
$\Delta_{B-V}$ and $M_V$ for blue galaxy selection fails to take
account of the youth (and thus brightness and blue-ness) of stellar
populations in galaxies at higher redshifts relative to their lower
redshift counterparts.  To counter this effect, \citet{andreon} and
\citet{loh} have advocated the use of differential $k$-corrections to
associate blue galaxies with the same spectral classes of galaxies at
all redshifts, and to take into account the expected luminosity
evolution of galaxies when defining the $M_V$ (or better still $M_K$)
limits.  The above studies all conclude that once these selection
biases have been eliminated the redshift evolution of $f_b$ is
significantly reduced, suggesting that there has been little evolution
in the cluster galaxy population since $z{\simeq}1$.  Similarly, when
using mass-selected samples ($\mathcal{M}{>}4{\times}10^{10}\Msol$)
\citet{holden} and \citet{van} find that the morphological composition
of clusters and the morphology-density relation has remained largely
unchanged since $z{\sim}0.8$, as opposed to luminosity-selected
($M_V<M_V^\star+1$) samples where significant redshift evolution is
seen \citep{smith05a,postman,desai}.

Another more subtle bias arises from the selection of the clusters
themselves.  BO84's cluster sample was a heterogeneous mixture of
clusters identified from photographic plates or by their association
with radio galaxies, and hence favored the inclusion of more extreme
clusters at high-redshifts, in particular those with higher blue
fractions as they would be easier to detect and identify.  Indeed
\citet{newberry} and \citet{andreon99} showed that the high-redshift
clusters of the BO84 sample are much more X-ray luminous and have
higher velocity-dispersions and central surface densities than their
low-redshift counterparts, suggesting that these selection biases
could mimic evolutionary effects, and illustrating the need for
well-defined samples of clusters spanning a large redshift range.
More recent studies of X-ray selected clusters again find that both
the scatter in the blue galaxy fraction and the trend with redshift is
much reduced \citep{smail,margoniner,ellingson,fairley,andreon} relative to 
those measured from optically-selected cluster samples
\citep[BO84,][]{depropris,goto}.  These latter samples contain a large
number of poor clusters, and although the influence of X-ray
luminosity ($L_X$), velocity dispersion ($\sigma$) or cluster richness
on $f_b$ is still much debated
\citep{margoniner01,wake,goto,popesso,aguerri}, there seems a general
consensus that for massive ($\sigma{>}600\kms$), X-ray luminous
clusters there is little scatter or evolution in $f_b$ out to
$z{\sim}0.8$ \citep{smail,homeier,poggianti,aguerri}.

Finally, the scatter in the blue galaxy fraction among the
intermediate-redshift ($0.15{\leq}z{\leq}0.3$) clusters is large;
values for individual clusters lying in the range $f_b\sim0-0.2$,
i.e.\ values typical of clusters at $z{=}0$ and $z{=}0.5$
respectively.  This has often been broadly attributed to the dynamical
status of the clusters, with actively merging clusters showing higher
fractions of blue or star-forming galaxies than those apparently
relaxed \citep{miller03,miller05}.  However, equally this could be due
to field galaxy contamination, as the blue galaxy fractions are
estimated by statistically subtracting foreground and background
objects by comparison to control fields, a process that becomes
increasingly uncertain at higher redshifts as the level of field
contamination rises.  However, in an analysis based on only
spectroscopically confirmed members of 60 clusters at $z<0.11$ covered
by the 2dFGRS, \citet{depropris} obtained a similarly large scatter in
the blue galaxy fraction, including some clusters with $f_b>0.4$, and
also found no correlations between $f_b$ and other cluster properties
including richness, substructure, concentration.

The sensitivity of \emph{ISO} and more recently \emph{Spitzer} at
mid-infrared (MIR) wavelengths has opened up a new window for studying
star-formation in galaxy clusters. The most straightforward
interpretation \citep{kennicutt07} of the 24$\mic$ emission is that
it traces the dust obscured star-formation, while the observed UV or
H$\alpha$ emission traces the unobscured one \citep{calzetti}.  MIR
observations of clusters have revealed a population of dusty
star-forming cluster galaxies
\citep[e.g.][]{fadda00,duc,biviano,geach,marcillac,bai,dressler08,haines08,saintonge}.
Indeed, these MIR-detected star forming galaxies often have optical
colors consistent with the passively evolving early-type cluster
galaxies \citep{wolf05,haines08} and would therefore be missed by the
traditional BO studies.  \citet{saintonge} therefore combined the
dusty star-forming cluster members, defined here as having
mid-infrared star-formation rates higher than $5\,\Msolpyr$, with the
traditionally selected blue cluster members -- all found within
$R\le1\Mpc$ of the cluster centers -- to show that the optical and IR
populations contribute roughly equally to the observed BO effect out
to $z\simeq0.8$.  However several studies have shown that the dusty
galaxies contribute significantly more to the integrated cluster star
formation rate than the optically selected blue galaxies.  For example
star formation rates derived from the [{\sc oii}] emission line are
typically $\sim10-30\times$ lower than rates estimated from IR
luminosities \citep[e.g.][]{metcalfe,geach09}. However, these
  previous studies have analyzed either single clusters or at most a
  handful of heterogeneously selected clusters over a large redshift
  range, with 1--2 clusters per redshift slice, such that no
  statistical analysis of trends with cluster properties or redshift
  has yet been done in the MIR.

In this article we revisit the BO effect, taking advantage of recently
obtained panoramic {\em Spitzer} MIR and ground-based NIR imaging of
22 clusters at $0.15{\le}z{\le}0.3$ from the Local Cluster Substructure
Survey (LoCuSS; http://www.sr.bham.ac.uk/locuss; see also
\S\ref{sec:locuss} for more details) plus comparable data for Coma and
Abell 1367 at $z{=}0.023$, five clusters from the Shapley supercluster
at $z{=}0.048$ covered by ACCESS (A Complete CEnsus of Star-formation
and nuclear activity in the Shapley supercluster), and Cl\,0024+17 at
$z{=}0.394$.  In contrast to the only previous MIR BO study
\citep{saintonge}, our large sample of 30 clusters in total allows us
to measure the scatter in the $\fsf$--redshift relation as a function of
redshift, and to explore the relationship between the scatter and
simple indicators of the dynamical state of the clusters.  We also use
the very wide field of view of our data to explore the radial
dependence of the MIR BO effect out to cluster-centric radii of
$R{\gs}2\Mpc$.  In summary, we present a statistical analysis of 30 clusters, aiming to
probe the balance between (i) the evolution of field galaxies that
fall into clusters, and (ii) the physical processes (some of which may
be related to cluster-cluster mergers) at play within the clusters, in
shaping the population of actively star-forming cluster galaxies.

In \S\ref{sec:data} we summarize the data used in this paper, and the
photometric analysis.  The main results are then presented in
\S\ref{sec:results}, followed by a summary and discussion in
\S\ref{sec:discuss}.  Throughout we assume \mbox{$\Omega_M{=}0.3$},
\mbox{$\Omega_\Lambda{=}0.7$} and \mbox{${\rm
    H}_0{=}70\,\kms\Mpc^{-1}$}.

\section{Data}
\label{sec:data}

We have assembled a dataset on 30 clusters at $0.02{\le}z{\le}0.4$
with NIR imaging extending out to the infall regions and
panoramic {\em Spitzer}/MIPS $24\mic$ photometry covering the same
regions.  Observational details for the clusters are listed in
Table~1.  New observations of 27 clusters at $z<0.3$ are described in
\S\S\ref{sec:locuss}~\&~\ref{sec:lowz} below.  The data and photometry
for the one remaining cluster at $z=0.4$ (Cl\,0024) were described in
detail by \cite{moran07} and \cite{geach}; in this paper we make use
of the Cl\,0024 master catalog\footnote{available from
http://www.astro.caltech.edu/$\sim$smm/clusters} and refer readers to the
relevant papers for further details.

\begin{table}
\centering
\begin{minipage}{80mm}
\caption{Observational details of the cluster sample}
\label{tab:sample}
\begin{tabular}{lccccc} \hline Cluster & z &
  NIR\footnote{Source of near-infrared data: (1) GOLDmine
    database \citep{gavazzi}; (2) WFCAM $K$-band data from ACCESS; (3)
    WFCAM $JK$-band data from LoCuSS; (4) NEWFIRM $JK$-band data from
    LoCuSS; (5) UKIDSS; (6) Treu et al. (2003)}
  &$r_{500}$& N$_{\rm gals}$ & $\fsf$\\ 
  name & & & (Mpc) & (${<}K^{*}\!{+}1.5$)\hspace{-2mm} & \\
  \hline 
  Abell 1367 & 0.022 & 1 & 0.83 &  47 & $0.068_{-0.039}^{+0.062}$ \\
  Coma       & 0.023 & 1 & 1.50 & 126 & $0.029_{-0.017}^{+0.025}$ \\
  Abell 3556 & 0.048 & 2 & 0.70 &  16 & $0.000_{-0.000}^{+0.109}$ \\
  Abell 3558 & 0.048 & 2 & 1.29 & 181 & $0.017_{-0.011}^{+0.017}$ \\
  Abell 3562 & 0.048 & 2 & 0.91 &  61 & $0.052_{-0.029}^{+0.047}$ \\
  SC\,1327-313&0.048 & 2 & 0.91 & 128 & $0.031_{-0.020}^{+0.028}$ \\
  SC\,1329-317&0.048 & 2 & 0.76 &  67 & $0.051_{-0.033}^{+0.039}$ \\ \hline
  RXJ1720.1+2638&0.160&3 & 1.53 & 183 & $0.048_{-0.016}^{+0.022}$ \\
  Abell 586  & 0.171 & 3 & 1.15 & 225 & $0.033_{-0.017}^{+0.021}$ \\
  Abell 1914 & 0.171 & 3 & 1.56 & 210 & $0.028_{-0.014}^{+0.019}$ \\
  Abell 2218 & 0.174 & 4 & 1.26 & 219 & $0.032_{-0.016}^{+0.020}$ \\
  Abell 2345 & 0.176 & 3 & 1.05 & 118 & $0.021_{-0.015}^{+0.025}$ \\
  Abell 665  & 0.182 & 4 & 1.38 & 233 & $0.049_{-0.019}^{+0.023}$ \\
  Abell 1689 & 0.182 & 3 & 1.50 & 259 & $0.043_{-0.014}^{+0.018}$ \\
  Z1883 / ZwCl\,0839.9+2937 & 0.194 & 3 & 1.11 &  94 & $0.095_{-0.034}^{+0.044}$ \\
  Z1693 / ZwCl\,0823.2+0425 & 0.223 & 3 & 1.00 & 110 & $0.031_{-0.022}^{+0.034}$ \\
  Abell 2219 & 0.225 & 4 & 1.49 & 363 & $0.043_{-0.013}^{+0.016}$ \\
  Abell 1763 & 0.228 & 3 & 1.22 & 262 & $0.036_{-0.016}^{+0.020}$ \\
  Abell 2390 & 0.230 & 4 & 1.50 & 303 & $0.073_{-0.024}^{+0.027}$ \\
  RXJ2129.6+0005&0.234&3 & 1.23 & 238 & $0.100_{-0.028}^{+0.032}$ \\
  Z2089 / ZwCl\,0857.9+2107 & 0.235 & 4 & 1.02 &  57 & $0.089_{-0.056}^{+0.076}$ \\
  Abell 1835 & 0.252 & 3 & 1.59 & 392 & $0.063_{-0.020}^{+0.022}$ \\
  Z348 / ZwCl\,0104.4+0048 & 0.254 & 5 & 1.00 & 131 & $0.129_{-0.044}^{+0.048}$ \\
  Z7160 / ZwCl\,1454.8+2233 & 0.258 & 3 & 1.13 & 114 & $0.049_{-0.023}^{+0.036}$ \\
  Abell 1758S& 0.273 & 3 & 1.38 & 189 & $0.052_{-0.023}^{+0.029}$ \\
  Abell 1758N& 0.279 & 3 & 1.16 & 313 & $0.053_{-0.018}^{+0.021}$ \\
  Abell 689  & 0.279 & 3 & 1.20 & 180 & $0.137_{-0.033}^{+0.039}$ \\
  Abell 697  & 0.283 & 3 & 1.51 & 286 & $0.076_{-0.022}^{+0.025}$ \\
  Abell 611  & 0.288 & 3 & 1.37 & 252 & $0.126_{-0.032}^{+0.025}$ \\ \hline
  Cl\,0024+17& 0.394 & 6 & 1.00 & 149 & $0.148_{-0.038}^{+0.043}$ \\ \hline
\end{tabular}
\end{minipage}
\end{table}

\subsection{LoCuSS}\label{sec:locuss}

LoCuSS is a multi-wavelength survey of a morphologically unbiased
sample of 100 X-ray luminous galaxy clusters at $0.15{\le}z{\le}0.3$
drawn from the ROSAT All Sky Survey cluster catalogs
\citep{ebeling98,ebeling00,bohringer}.  The overall aim is to
constrain the cluster-to-cluster scatter in the observable properties
(e.g.\ X-ray temperature, integrated SZ-effect $Y$-parameter, star-formation
rate, far-infrared galaxy luminosity function) of massive clusters at
low redshift, and to interpret these observables in the
context of hierarchical assembly, aided by gravitational lensing probes
of the distribution of dark matter in the clusters
\citep[e.g.][]{smith08}.  These analyses will, for example, deliver
new constraints both on the normalization, shape and scatter of
mass-observable scaling relations required for precision cluster
cosmology, and on the evolutionary pathways of gas-rich field galaxies
into passive cluster early-type galaxies.  Early results on these two
complementary aspects of the survey can be found in \cite{zhang08}, \citet{marrone} and
\cite{haines09}. 

 The first batch of 30 clusters in our survey
benefits from a particularly rich dataset, including:
Subaru/Suprime-Cam optical imaging \citep{okabe09}, {\em Spitzer}/MIPS
$24\mic$ maps, GALEX near- and far-ultraviolet (NUV/FUV) imaging, and
near-infrared (NIR; $J,K$) imaging from UKIRT/WFCAM and
KPNO-4m/NEWFIRM.  All of these data embrace at least a half-degree
field of view centered on each cluster, and thus probe the clusters
out to ${\sim}1-2$ virial radii.  We have also been awarded
500\,ksec on {\em Herschel} as an Open Time Key Programme to observe
this sample at $100$ and $160\mic$ with PACS.  These 30 clusters were
selected from the parent sample simply on the basis of being
observable by Subaru on the nights allocated to us. In principle these 30 
should therefore not suffer any gross biases towards one type of cluster or
another (e.g.\ cool core cluster, merging cluster, etc.).  In this
paper we analyze the 22 clusters from the full sample of 30 for which both
NIR and MIR data are in-hand.

\subsubsection{LoCuSS Mid-IR Observations}

Each cluster was observed across a $25^\prime{\times}25^\prime$ field
of view at $24\mic$ with MIPS \citep{rieke} on board the {\em Spitzer Space
  Telescope}\footnote{This work is based in part on observations made
  with the Spitzer Space Telescope, which is operated by the Jet
  Propulsion Laboratory, California Institute of Technology under a
  contract with NASA (contract 1407).}  \citep{werner}, consisting of
a $5{\times}5$ grid of MIPS pointings in fixed cluster or raster mode (PID: 40872; PI: G.P. Smith).
At each grid point we performed two cycles of the small-field
photometry observations with a frame time of 3s, producing a total per pixel
exposure time of 90s.  The central $5^\prime{\times}5^\prime$ tile of
some clusters had already been imaged by Guaranteed Time Observations
program 83 to a much deeper depth (${\sim}3000\,{\rm s/pixel}$); these
data were combined with our $24\mic$ data where available to give
complete coverage of the entire $25^\prime{\times}25^\prime$ field
centered on each cluster in the sample.

The $24\mic$ data were reduced and combined with the Data Analysis
Tool (DAT) developed by the MIPS instrument team \citep{gordon}.  A
few additional processing steps were also applied as described in
\citet{egami}.  The data were resampled and mosaicked with half of the
original instrument pixel scale (1\farcs245) to improve the spatial
resolution.  The $24\mic$ mosaics were analyzed with SExtractor
\citep{bertin}; following SWIRE \citep[Spitzer Wide-area Infra-Red
Extragalactic Legacy Survey;][]{swire} we estimated the flux of
objects within an aperture of diameter $21''$, and applied an aperture
correction factor of $1.29$.  The flux detection limits and completeness of
each mosaic were determined by individually inserting 500 simulated
sources for a range of fluxes and determining their detection rate and
recovered fluxes, using identical extraction procedures.  The sources
used in the simulations were formed by extracting isolated, high
signal-to-noise and unresolved sources from the mosaic itself. From
these simulations, we estimate that the 90\% completeness limits of
our $24\mic$ mosaics are typically 400$\mu$Jy, and for individual clusters in the range 300--500$\mu$Jy, the variation due to changes in the background cirrus level.

\subsubsection{LoCuSS Near-IR Observations}

The same 22 clusters were observed either with WFCAM \citep{wfcam} on
the 3.8-m United Kingdom Infrared Telescope (UKIRT)\footnote{UKIRT is
  operated by the Joint Astronomy Centre on behalf of the Science and
  Technology Facilities Council of the United Kingdom.} in
March--November 2008 
or with NEWFIRM on the 4.0-m Mayall telescope at Kitt Peak\footnote{Kitt
  Peak National Observatory, National Optical Astronomy Observatory,
  which is operated by the Association of Universities for Research in
  Astronomy (AURA) under cooperative agreement with the National
  Science Foundation.} on 16--18 May 2008. 
 The WFCAM data were obtained using the same observing
strategy as used by the UKIDSS Deep Extragalactic Survey
\citep{ukidss}, covering $52'{\times}52'$ to depths of $K{\sim}19$,
$J{\sim}21$, with exposure times of 640s, pixel size of $0.2''$ and
${\rm FWHMs}{\sim}0$.8--1.$1''$.  The NEWFIRM data consist of dithered and
stacked $J$- and $K$-band images with exposure times of 1800s, and
${\rm FWHM}{\sim}1$.0--1$.5''$, covering the instrument field of view
$27'{\times}27'$ with a $0.4''$ pixel-scale.  Each individual exposure
was astrometrically and photometrically calibrated using 2MASS stars
in the field, and stacked using {\sc iraf}, producing mosaics also
reaching depths of $K{\sim}19$, $J{\sim}21$.

\begin{figure}
\plotone{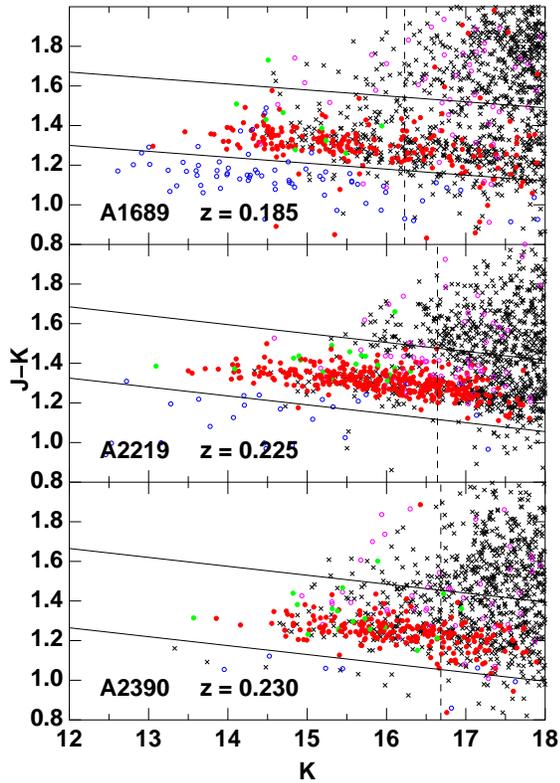}
\caption{$J-K/K$ color-magnitude diagrams for galaxies within
  $1.5r_{500}$ of clusters A\,1689, A\,2219 and A\,2390.  These
  three clusters are shown because their spectroscopic redshift
  catalogs are the most complete among the 22 clusters from LoCuSS.
  Filled symbols indicate spectroscopically-confirmed cluster members, 
 with green (red) colors indicating those (not) having $L_{IR}{>}5{\times}10^{10}L_{\odot}$ based on their $24\mic$ fluxes.  Blue and
  magenta open symbols indicate foreground and background galaxies
  with redshifts respectively, while crosses indicate those galaxies
  without redshift information.  The pair of sloping lines indicate
  the upper and lower boundaries of the color-magnitude selection
  used to identify probable cluster members.  The dashed line
  indicates our faint magnitude limit of $\Kstar+1.5$.  }
\label{cm}
\end{figure}

Following \citet{haines09}, probable cluster members were identified
from the $JK$ photometry (with $J{-}K$ colors determined in $2''$
diameter apertures), based on the empirical observation that galaxies
of a particular redshift lie along a single narrow $J{-}K/K$
color-magnitude relation, as shown in Fig.~\ref{cm} for A\,1689,
A\,2219 and A\,2390.  This relation evolves redward monotonically with
redshift to $z{\simeq}0.5$ \citep[see][]{haines09}.  The NIR colors
of galaxies are relatively insensitive to star-formation history and
dust extinction, with the $J{-}K$ color varying by only ${\sim}0.1{\rm mag}$
across the entire Hubble sequence, and hence only a single sequence is
seen, unlike in the optical where separate red and blue sequences are
visible.  This is demonstrated in Fig.~\ref{cm} by the
$24\mic$-detected cluster members (green symbols) showing the same
$J{-}K$ colors as those cluster members not detected at $24\mic$,
albeit with more scatter.  Indeed the $24\mic$ sources often lie well
above the sequence, presumably reddened by dust, necessitating the use
of a selection box that is asymmetric around the ``red sequence'' of
cluster members.

The NIR color cuts are particularly efficient at removing foreground
galaxies, with ${>}80\%$ of $K{<}\Kstar{+}1.5$ galaxies
spectroscopically confirmed as having
$z_{sp}{-}z_{cluster}{<}-0.05(1+z)$ (shown by blue open symbols in
Fig.~\ref{cm}) excluded by the lower $J-K$ color cut.  However there
remains significant contamination by background galaxies in the range
$z_{cluster}{<}z{\la}0.5$ (open magenta symbols), driven partly by the
red envelope of our color-magnitude selection function in an attempt
to minimize the loss of heavily reddened cluster members (see upper
sloping line in each panel of Fig.~\ref{cm}).  We correct statistically
for this contamination, using galaxy counts satisfying the same
color-magnitude selection in two control fields.  We analyzed the
UKIDSS-DXS Lockman Hole and XMM-LSS fields \citep{ukidss}, based on
$JK$ photometry covering $2.1\,{\rm deg}^2$ in total, obtained via the same
observing strategy, reduction and calibration pipelines, as our WFCAM
data.  These fields also have publicly available $24\mic$ MIPS
photometry from SWIRE DR2 catalogs \citep{swire}\footnote{The SWIRE
  DR2 catalogs are available here:
  http://swire.ipac.caltech.edu/swire/astronomers/data\_access.html}
which are complete to $450\uJy$.

With the exception of the clusters highlighted in Fig.~\ref{cm}, the
spectroscopic redshift information on galaxies in clusters from the
LoCuSS sample is currently very sparse.  For the purposes of this
paper we therefore define galaxies as being cluster members if they
satisfy the color-cuts discussed above.  The fraction of star-forming
galaxies in each cluster, $\fsf$, calculated in
\S\ref{sec:results} from these cluster galaxy catalogs are therefore
net of the statistical field subtraction described above, and the
error bars include a term to account for the uncertainties in this
subtraction.

\subsection{Low-redshift subsample}\label{sec:lowz}

For Coma and Abell 1367 we used archival $24\mic$ {\em Spitzer}/MIPS
data covering $2{\times}2\,{\rm deg}^2$ in the case of Coma (PID: 83, PI
G. Rieke) and $30^\prime{\times}30^\prime$ for Abell 1367 (PID: 25, PI
G. Fazio). Both datasets were obtained in scan mode, and are complete to
${\sim}330\uJy$, corresponding to $L_{IR}{\sim}1.4{\times}10^8L_\odot$
\citep{bai06}.  Note that the $24\mic$ mosaics do not extend out to
the aperture of $1.5r_{500}$ adopted in this paper, we therefore
estimate the $24\mic$ fluxes of galaxies that lie outside the observed
$24\mic$ field of view based on their IRAS $60\mic$ fluxes and the
empirical relation $f_{60}/f_{24}{\sim}10$ \citep{soifer}.  The IRAS
Faint Source Catalogue $60\mic$ completeness limit is 0.6\,Jy
\citep{beichman}, corresponding to $L_{IR}\sim4{\times}10^{10}L_\odot$
for Coma and A\,1367, or just below our $L_{IR}$ selection limit used in \S\ref{sec:results}. $K$-band photometry and redshifts for both of
these clusters were taken from the GOLDMine\footnote{available at
  http://goldmine.mib.infn.it/} database \citep{gavazzi}.

For the five clusters forming the core of the Shapley supercluster
(A\,3556, A\,3558, A\,3562, SC\,1329-313 and SC\,1327-312) at
$z{=}0.048$, we use unpublished {\em Spitzer} $24\mic$ MIPS imaging
(PID: 50510, PI: C.\ Haines), UKIRT/WFCAM $K$-band photometry 
and redshifts from the ACCESS survey 
all of which cover a ${\sim}2.5{\times}1.5\,{\rm deg}^2$ field of
view.  The $24\mic$ mosaics are complete to ${\sim}400\uJy$,
corresponding to $L_{IR}{\sim}1{\times}10^{9}L_\odot$ at the
supercluster redshift.  The WFCAM $K$-band data has exposure times
300s, pixel-scale $0.2''$ and is $90\%$ complete to $K{\sim}17.5$
($\Kstar{+}5.5$), while our redshift coverage is ${>}95\%$ complete
to $\Kstar{+}1.5$.  At these low redshifts many of the $24\mic$
sources are resolved, and so for those galaxies with NIR diameters 
greater than our standard $21''$ MIR aperture, 
we used a series of circular
apertures of diameter up to $2'$ ($45''$ for Shapley), and
the optimal diameter chosen to encircle all of the galaxy's NIR flux.

\subsection{Bolometric Infrared Luminosities of $24\mic$ Sources}

\begin{figure*}
\plotone{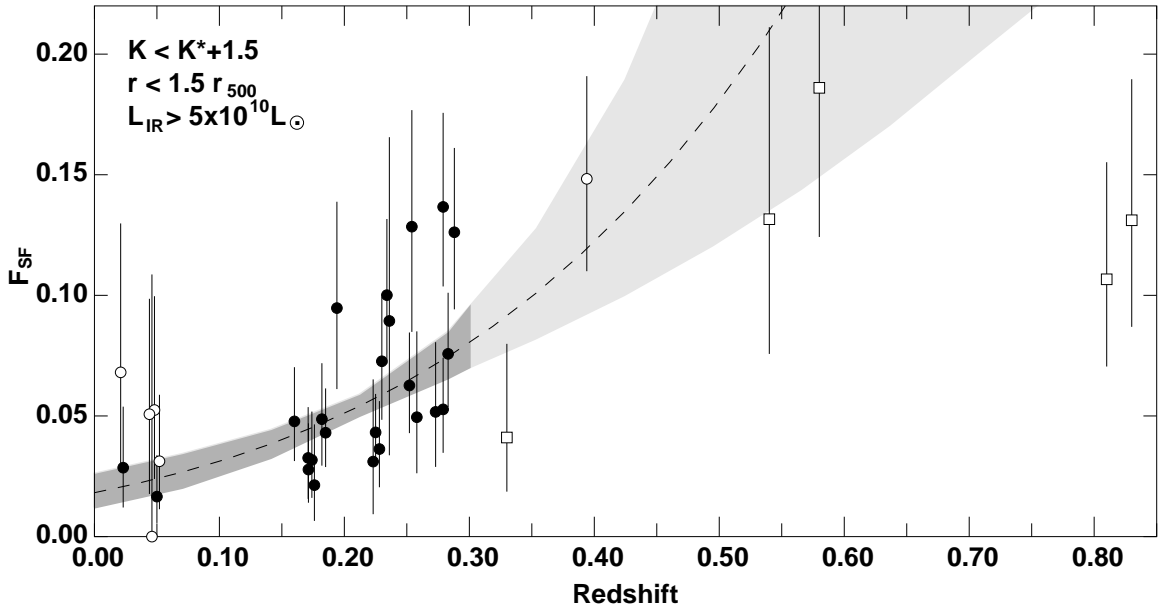}
\caption{The mid-infrared Butcher-Oemler effect.  The estimated
  fraction of $M_K{\le}M_K^{\star}{+}1.5$ cluster members within
  $1.5r_{500}$ having $L_{IR}{>}5{\times}10^{10}L_\odot$.  The dashed line
  indicates the best-fitting evolutionary fit of the form
  $\fsf{=}f_{0}(1+z)^n$ to the $z{<}0.3$ clusters, and the shaded region
  indicates the $1\sigma$ confidence region to the fit, with the lighter colors showing the extrapolation of the fit beyond $z{=}0.3$.
 Solid circles indicate clusters with $L_{X}{>}3{\times}10^{44}\ergs$, while open circles indicate less X-ray luminous clusters. The open squares indicate the values of $f_{SF}$ taken from the \citet{saintonge} $z{>}0.3$ sample.} 

\label{bo}
\end{figure*}

The bolometric luminosity, $L_{IR}$, of each $24\mic$-selected cluster
galaxy is estimated from its $24\mic$ flux using a range of infrared
SEDs \citep{chary,chanial,dale,lag03,lag04}, as a function of redshift
following \citet{lefloch}.  We classify as star-forming those
  galaxies with $L_{IR}{>}5{\times}10^{10}L_\odot$, which corresponds
  to the $24\mic$ completeness limits for our five highest redshift
  clusters ($0.27{<}z{<}0.29$). Note that for all the remaining 
  clusters at $z{\simeq}0.2$, this $L_{IR}$ limit lies in the range 550--1500$\mu$Jy,
  and hence is always well above the corresponding completeness limit
  of the cluster.

Assuming that the 24$\mic$ emission is due entirely to star-formation,
this limit corresponds to a star formation rate (SFR) of
${\sim}8\Msolpyr$ \citep{kennicutt}.  The choice of infrared SEDs
introduces a spread of $L_{IR}/S_{24\mic}$ ratios, due to the range of
dust temperatures assumed.  This spread varies systematically with
redshift, and is of the order $0.1-0.2\,{\rm dex}$ over $0{<}z{<}0.3$.
These systematic uncertainties are incorporated into the error bars quoted in
\S\ref{sec:results}.  More accurate estimates of $L_{IR}$ will be
possible in future with the availability of {\em Herschel} $100\mic$
and $160\mic$ photometry, which will allow us to fully model the
reprocessed thermal emission from the dust.

Some of the 24$\mu$m emission could be due to dust-enshrouded active
  galactic nuclei (AGN), as a significant fraction of infrared
  galaxies are known to be powered by a combination of AGN activity
  and star-formation. 
  At the typical infrared luminosities of our sample
  (0.5--2${\times}10^{11}L_{\odot}$), the primary power source of the
  24$\mu$m emission has been shown to be star-formation in ${\ga}90$\%
  of the cases \citep[e.g.][]{goulding}. There is also a strong
  almost-linear correlation between the 24$\mu$m and
  extinction-corrected Pa$\alpha$ fluxes over four decades in
  luminosity among local infrared galaxies, the latter emission shown
  from HST/NICMOS imaging to be in the form of nuclear star-formation
  rings or more extended emission from spiral arms or disks
  \citep{alonso}.

  For our low-redshift cluster galaxy sample, we can constrain the
  dominant power source behind the observed 24$\mu$m emission, using
  both optical spectroscopy and the morphology of the 24$\mu$m emission
  (extended/unresolved). For this purpose we use SDSS DR7 \citep{sdssdr7} for Coma
  and Abell 1367 members, or \citet{smith07} for galaxies in the
  Shapley supercluster, to place 24$\mu$m-selected galaxies on the emission-line diagnostic diagrams of
  \citet{baldwin}.  We only consider it likely that an AGN as
  the dominant contributor to the infrared flux if a galaxy is {\em
    both} optically-classified as an AGN {\em and} the 24$\mu$m
  emission is unresolved, as \citet{brand} found in a study of
  $0.15{<}z{<}0.30$ galaxies with $f_{24}{>}0.3$mJy that the bulk of those
  galaxies classified optically as AGN also had infrared colors
  indicative of PAH emission from star-formation. 

 We find that all 13 infrared-luminous
  galaxies in Coma/A1367 have {\em
    either} extended 24$\mu$m emission or optical emission-line ratios
  indicative of star-formation. In Shapley supercluster, we
  find just three of 29 infrared-luminous galaxies are classified as an AGN.

  Considering also X-ray bright AGN, \citet{martini} studied X-ray
  luminous ($L_{X}{>}10^{41}$erg\,s$^{-1}$) AGN in eight clusters at
  $0.06{<}z{<}0.31$, obtaining an AGN fraction of $\sim5$\% among
  $M_{R}{<}-20$ galaxies. \citet{eastman} found
  evidence that the X-ray AGN ($L_{X}{>}10^{42}$erg\,s$^{-1}$) fraction increases by a factor 20 from $z{=}0.2$ to $z{=}0.6$. We consider what fraction of these X-ray
  luminous AGN are also strong MIR emitters. For Abell
  1689, which is in the \citet{martini} sample, only one of our
  infrared-bright sources is found to be an X-ray AGN, while for Abell
  1758, we find just two of our 82 24$\mu$m detections to coincide
  with an X-ray point source \citep{haines09}.

\section{Results}
\label{sec:results}

\subsection{The Mid-Infrared Butcher-Oemler effect}\label{sec:bo}

For each cluster, we compute the fraction of star-forming cluster
members ($\fsf$) within $1.5r_{500}$
\citep[${\sim}1.0r_{200}$;][]{sanderson03}, after selecting by
rest-frame $K$-band magnitude ($M_K{\le}M_K^\star{+}1.5$), and
bolometric infrared luminosity ($L_{IR}{>}5{\times}10^{10}L_\odot$).  We
exclude the brightest cluster galaxy (BCG) from each cluster, due to
their unique star-formation histories \citep{lin} and the direct
link between BCG activity and the presence of cooling flows within
clusters \citep{edge}.
 We
model the redshift evolution of $\Kstar$ as a \citet{bc03} stellar
population formed at $z{=}4$ with an exponentially-decaying SFR of
time-scale 1\,Gyr, normalized to match the present-day value of 
$M_K^\star{=}-24.60{\pm}0.03$ \citep{jones} for field galaxies. We note that this value is slightly brighter than those observed for galaxy clusters, e.g. $M_K^\star{=}-24.58{\pm}0.40$ \citep{balogh01}, $M_K^\star{=}-24.34{\pm}0.01$ \citep{lin04} (for $\alpha{=}1.1$), but we prefer to refer to a ``global'' value for $M_{K}^{*}$.
 The measurements of
$r_{500}$ for clusters at $0.15{<}z{<}0.3$ are obtained from analysis of
Chandra X-ray observations, both from the archive and our own Cycle~10
observations (PID: 10800565; PI: G.P. Smith); the details of this analysis can be found in \citet{sanderson09}.  Values of $r_{500}$
for the remaining clusters are taken from the literature \citep{sanderson}.

The value of $\fsf$ for each cluster is listed in
Table~\ref{tab:sample}; the quoted uncertainties include the binomial
error calculated using the formulae in \citet{gehrels}, and
uncertainties on the statistical subtraction of field galaxies.  We
show $\fsf$ versus redshift -- the MIR Butcher-Oemler effect -- in
Fig.~\ref{bo}, finding a steady increase in $\fsf$ with redshift,
from $\langle\fsf\rangle{=}0.035{\pm}0.023$ at $z{<}0.05$, to
$\langle\fsf\rangle{=}0.053{\pm}0.027$ at $0.15{\le}z{\le}0.25$,
and $\langle\fsf\rangle{=}0.085{\pm}0.038$ at
$0.25{\le}z{\le}0.29$, where the quoted uncertainties are the rms
scatter around the means.  This trend is consistent with that found
for the original BO study, whereby $f_{b}$ increases from
$0.053{\pm}0.054$ at $z{<}0.08$ to $0.098{\pm}0.062$ over
$0.17{\le}z{\le}0.28$, although we caution that given the rather diverse counting radii, cluster and galaxy selection criteria used, this may be coincidental.

To quantify the redshift evolution of $\fsf$ we fit the following
relation to the individual data points at $z<0.3$ shown in
Fig.~\ref{bo}: $\fsf{=}f_{0}(1+z)^n$.  The best-fit exponent is
$n{=}5.7^{+2.1}_{-1.8}$; the corresponding curve and $1\sigma$
confidence interval are over-plotted on Fig.~\ref{bo}.  Extrapolating
the best-fit relation (shown as lighter shaded region) reveals that it is consistent with the value of
$\fsf$ that we measure for Cl\,0024 at $z=0.395$.  Note that our
measurement of $\fsf{=}0.15{\pm}0.04$ for this cluster is consistent
with that obtained by \citet{geach}. 
 The rapid rate of evolution does not depend
strongly on the choice of limiting radius; repeating the calculation
using galaxies selected within $r_{500}$ we obtain
$n{=}5.0^{+2.9}_{-2.3}$, however the values of $\fsf$ are in this case  
systematically lower than their counterparts within $1.5r_{500}$ by
${\sim}20\%$. 

Our use of NIR color cuts could exclude some star-forming cluster
  members, whose colors in Fig.~\ref{cm} show a much greater spread
  than their passive counterparts, resulting in a fraction lying
  outside our color limits. To test this possibility, we redo the
  analysis, this time without any color cuts. The overall trend and
  scatter are unchanged, although as the correction for field
  contamination increases, so do the uncertainties both in $\fsf$ for
  individual clusters, and the overall measured evolutionary trend,
  for which we obtain $n{=}6.5^{+2.5}_{-2.0}$. The slight increase in
  $\langle\fsf\rangle{=}0.099{\pm}0.036$ at $0.25{\le}z{\le}0.29$ obtained
  when we remove the color cuts, suggests that a small fraction
  (${\sim}10$\%) of 24$\mu$m cluster members may be lost by our color
  cuts, but that this is not significant.

The measured level of redshift evolution is also consistent with the
MIR BO-study of eight clusters over $0.02{<}z{<}0.83$ by
\citet{saintonge}, at least for their $z{<}0.4$ clusters.  The absence
of a statistical sample of clusters at $z{>}0.3$ clusters in either this
study or that of \citet{saintonge}, makes it difficult to interpret
whether the strong redshift evolution might extend to higher redshift.
However their two clusters at $z{\simeq}0.8$ (shown as open squares in Fig.~\ref{bo}) are inconsistent with an extrapolation of our best-fit evolutionary model at 2$\sigma$. This suggests that the MIR BO effect might saturate beyond
$z{\sim}0.5$, although we note that they use a smaller fixed 1\,Mpc counting radius to estimate their $\fsf$.

A possible caveat to this analysis is that our $L_{IR}$ threshold for
selecting star-forming galaxies lies within the exponential region of
the infra-red luminosity function (LF), particularly at $z{\sim}0$
\citep[estimate $\Lirstar{=}3{\times}10^{10}L_{\odot}$ for
Coma]{bai06} so that small changes in the threshold could produce
large changes in $\fsf$.  We are sensitive here to only the most
actively star-forming galaxies, missing a significant fraction of the
population of normal star-forming galaxies, which are known to form a
well-defined sequence in the specific-SFR/stellar mass plane
\citep{noeske}.  As a result we might expect our values of $\fsf$ to
be systematically below the $f_b$ values obtained for the same
clusters.  However, when averaging over the six clusters in common
with \citet{smail}, we find no significant difference between our
$\fsf$ values and their $f_b$ values.  We also note that only two of
the low redshift clusters (Coma and A\,3558) have X-ray luminosities
comparable with the clusters from LoCuSS at $0.15\le z\le0.3$.
Strictly speaking this analysis therefore does not compare like with
like at low and intermediate redshifts.  However, the results in this
and subsequent sections are robust to the exclusion of all clusters at
$z<0.1$ except for Coma and A\,3358, in which case we now obtain $\fsf{=}7.8_{-2.3}^{+2.3}$ within $1.5\,r_{500}$. Indeed the primary driver of our observed rapid evolution in $f_{SF}$ is that produced {\em within} the LoCuSS sample, rather than the comparison to the low-redshift subsample.
This mis-match is therefore not a major concern, however it is unfortunate that there does not exist a suitable comparable set of panoramic mid-infrared observations of rich clusters at $z{<}0.1$.

\subsection{Mid-IR Butcher Oemler Effect and the Global Decline in
  Star-formation}
\label{evol}

\begin{figure}
\plotone{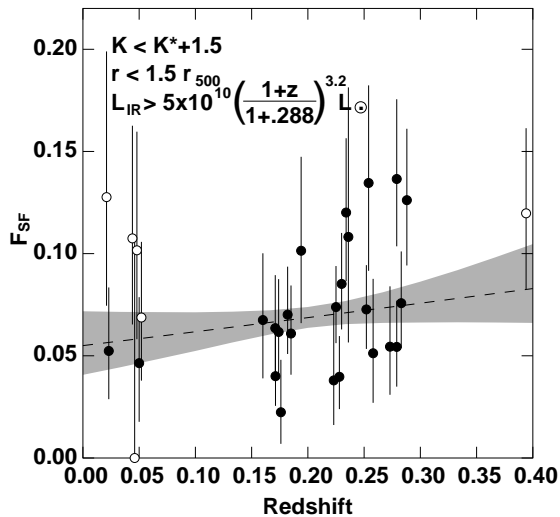}
\caption{As in Fig.~\ref{bo}, but taking into account the evolution of
  the mid-infrared luminosity function of field galaxies,
  $\Lirstar{\propto}(1+z)^{3.2}$, when identifying galaxies as
  star-forming.  Symbols are as in Fig.~\ref{bo}.}
\label{bo_evol}
\end{figure}

We now investigate whether the redshift evolution of $\fsf$ is caused
by physical processes operating within clusters, or simply reflects
the global decline in star formation since $z{\sim}1$, or a
combination of both.  The redshift evolution of $\fsf$ discussed above
is stronger than that of the global star formation rate.  For example,
\citet{zheng} showed that the specific UV+IR-determined SFR has
declined by a factor ${\sim}10$ since $z{\sim}1$, independent of galaxy
mass, while the specific SFRs of local galaxies are $\sim0.3\dex$ lower
than those of $0.2{<}z{<}0.4$ galaxies of the same mass.
\citet{lefloch} find strong evolution in the global infrared LF since
$z\sim1$, characterized by $\Lirstar{\propto}(1+z)^m$ with
$m{=}3.2^{+0.7}_{-0.2}$, similar to the $m=3.0\pm0.3$ obtained by
\citet{perez}.  \citet{noeske} also find that the median SFR at a
fixed stellar mass of the {\em entire} sequence of star-forming
galaxies shifts downwards by a factor of three from $z{=}0.98$ to
$z{=}0.36$, corresponding to an evolution of
$\sfr{\propto}(1+z)^{2.9}$. 

Studies of the MIR LFs of nearby clusters find the shape and $L^{*}_{IR}$ to be consistent with that of field galaxies, while the evolution with redshift to $z{\simeq}0.8$ of $L^{*}_{IR}{\propto}(1+z)^{3.2\pm0.7}$ is also indistinguishable from the field \citep{bai}.
 Similarly, \citet{finn} find that the H$\alpha$ luminosities of individual cluster galaxies have declined by a factor of up to ${\sim}10$ since $z{\sim}0.75$, comparable to that of field galaxies over a similar redshift interval.

To quantify the excess evolution of $\fsf$ over the global evolution
of star formation we therefore repeat the analysis in
\S\ref{sec:bo}, this time allowing the luminosity threshold above
which galaxies are counted in the numerator of $\fsf$ to scale with
redshift as follows: $L_{IR}{>}5{\times}10^{10}[(1+z)/(1+z_{\rm
  max})]^k$, with $k{=}3.2$ chosen to be representative of the results
discussed above, and $z_{\rm max}{=}0.288$, corresponding to the
highest cluster redshift to which we fit the redshift evolution model.
This is analogous to the use of differential $k$-corrections to
identify ``blue'' galaxies with the same spectral classes of galaxies
at all redshifts \citep[e.g.][]{andreon,loh}, rather than the fixed
$\Delta_{B-V}{=}0.2$ criterion of BO84.  We re-plot the MIR BO effect using
this selection function in Fig.~\ref{bo_evol}, and find (as expected)
that the redshift evolution has largely disappeared:
$\fsf{=}0.072{\pm}0.044$ for the seven clusters at $z{<}0.1$, and
$\fsf{=}0.074{\pm}0.033$ for the 22 clusters at $0.15{<}z{<}0.3$.
Again, we fit a model to the data of the form $\fsf(z)\propto(1+z)^n$,
obtaining $n{=}1.22^{+1.52}_{-1.38}$, as shown by the dashed curve and
error envelope in Fig.~\ref{bo_evol}. 

\begin{figure}
\plotone{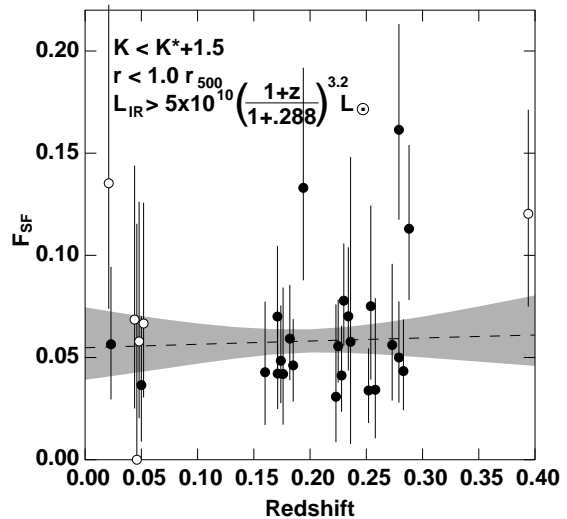}
\caption{As in Fig.~\ref{bo_evol}, but considering only those galaxies
  within $r_{500}$.  Symbols are as in Fig.~\ref{bo}.}
\label{bo_r500}
\end{figure}

This level of evolution is consistent with the expected increase in
the fraction of galaxies accreted by massive clusters within the
previous Gyr.  \citet{berrier} show that although the accretion rate
of galaxies into massive clusters has remained constant since
$z{\sim}1$, half of the cluster galaxy population has been accreted
since $z{\sim}0.4$, and so the {\em fraction} of recently accreted galaxies
should appear to double from $z{\sim}0$ to $z{\sim}0.4$.  This
suggests that the ``residual'' redshift evolution seen in
Fig.~\ref{bo_evol}, albeit at modest statistical significance, is
attributable to the fate of galaxies after they have fallen into
clusters.  Galaxy populations that have recently fallen into clusters
are most easily identified in photometric observational data at large
cluster centric radii, thus overcoming the projection of the
three-dimensional distribution of galaxies onto the sky.  To gain a
rough idea of the location of the galaxies responsible for the
redshift evolution seen in Fig.~\ref{bo_evol} we modify further the
selection function, this time restricting the range of cluster-centric
radii to ${<}r_{500}$, again using the ``differential $k$-correction''
approach to $L_{IR}$ selection.  The results of this modified
selection are shown in Fig.~\ref{bo_r500}, with the best-fit redshift
evolution model again shown as the dashed curve; the best fit model
has $n{=}0.32^{+1.74}_{-1.66}$.  We also fit a model with just a single
parameter -- a redshift-invariant value of $\fsf$ -- obtaining
$\fsf{=}0.056{\pm}0.004$ with a reduced chi-squared value of $0.72$,
confirming that the scatter of the data around this value can be
explained simply by the observational uncertainties, without recourse
to either intrinsic cluster-to-cluster scatter or redshift evolution.

In summary these results are consistent with the global decline in IR
activity in field galaxies since $z{=}1$.  As field galaxies fall into
clusters some of them suffer an increase in IR activity at ${>}r_{500}$,
presumably due to star formation induced by gentle processes in their
local environment at large radii.  The amplitude of this increase, as
measured via $\fsf$, evolves with redshift at a rate which is not significantly different from zero, but nonetheless consistent with
the fractional rate at which galaxies are accreted into rich clusters.
Nevertheless, there is still significant cluster-to-cluster scatter, and 
so now we turn to whether this variation in $\fsf$ 
might be due to some global property of the cluster.

\subsection{Correlations with cluster properties}

\begin{figure}
\plotone{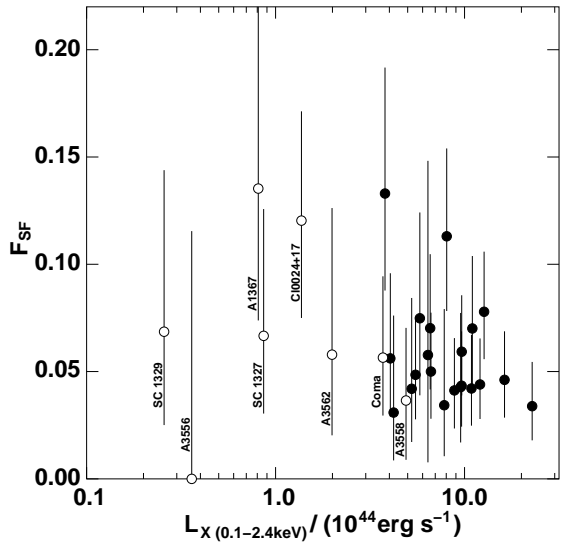}
\caption{Relation between the fraction of star-forming cluster
  galaxies within $r_{500}$ and the X-ray luminosity.  Solid symbols
  indicate LoCuSS clusters, while labelled open symbols indicate the
  non-LoCuSS subsample.}

\label{bo_lx}
\end{figure}

A number of previous studies \citep[e.g.][]{wake,popesso,aguerri} have
looked for correlations between the X-ray luminosity of clusters and
the fraction of star-forming galaxies (or equivalently
blue/emission-line), with the aim of ascertaining if ram-pressure
stripping and/or starvation could be the dominant mechanisms for
quenching star-formation in infalling galaxies, and driving the
observed SF-density relation.

In Fig.~\ref{bo_lx} we plot the fraction of star-forming cluster
galaxies within $r_{500}$ against the cluster X-ray luminosity in the
0.1--2.4\,keV band taken directly from the ROSAT All Sky Survey
cluster catalogs \citep{ebeling98,ebeling00,bohringer}, using the same evolving $L_{IR}$ cut as described in \S\ref{evol}. 
We exclude here Abell 689 as its X-ray luminosity is dominated by a BL Lac. As
previously mentioned, clusters in the LoCuSS sample are more X-ray
luminous ($L_{X}{>}2{\times}10^{44}\ergs$) than all of the other
clusters except Coma and A\,3558.  Overall there is no apparent trend
of $f_{SF}$ with X-ray luminosity, in agreement with \citet{wake}, which might
seem to rule out ICM-related processes.  However, this may simply be a
saturation effect, in that ram-pressure stripping is effective at
stripping the gas in all infalling galaxies even for the lowest X-ray
luminosity clusters ($L_{X}{\sim}3{\times}10^{43}\ergs$) in our
sample.  Indeed, \citet{poggianti} find a strong anti-correlation
between the fraction of emission-line galaxies and cluster velocity
dispersion for $\sigma{\la}550\kms$, but for richer systems there are
no systematic trends.  This may also explain the observed negative
trend between $f_b$ and $L_X$ seen by \citet{popesso}, as their
cluster sample extends to much poorer systems than ours or that of
\citet{wake}.

\begin{figure}
\plotone{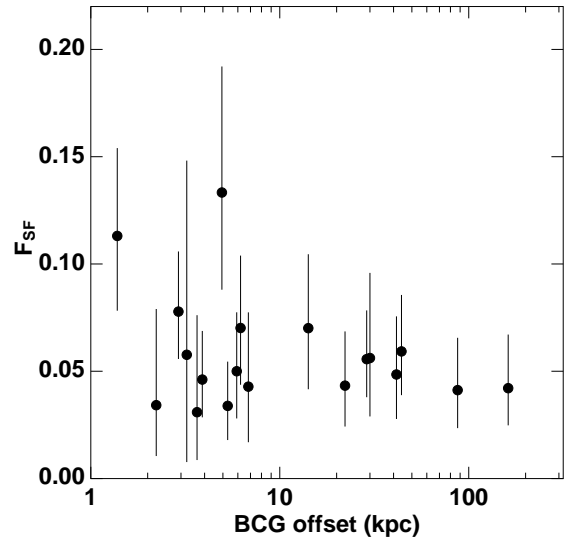}
\caption{Relation between the fraction of star-forming cluster
  galaxies within $r_{500}$ and the offset of the brightest cluster
  galaxy from the peak of X-ray emission.}
\label{bo_bcg}
\end{figure}

As discussed in \S\ref{intro} it has been suggested that the level of
star-formation activity in a galaxy cluster correlates with its
dynamical status, with merging clusters showing increased activity
with respect to their undisturbed counterparts
\citep{miller03,miller05,metcalfe}. No single measure exists in the literature
that unambiguously identifies a cluster as being ``disturbed'', i.e.\
undergoing a merger.  Nevertheless, several measures do appear to
correlate with cluster dynamical status, such as the presence or
absence of a cool core in the X-ray temperature profile, the cuspiness
of the density profile, or the offset of the brightest cluster galaxy
(BCG) from the peak of X-ray emission \citep[e.g.][]{smith05b}.  In
these cases merging clusters are typically identified with non-cool
core clusters with flatter density profiles, and BCGs with large
offsets from the X-ray peak.

In Fig.~\ref{bo_bcg} we therefore plot the fraction of star-forming
galaxies within $r_{500}$ versus the projected physical distance
between the BCG and the peak of X-ray emission, for the LoCuSS
subsample, the latter measurement being taken from Sanderson et al.\
(2009).  Interestingly there is no obvious correlation
between the BCG offset and $\fsf$, or alternatively when 
comparing the cuspiness of the
cluster density profiles with $\fsf$.  We also obtain similar results
when considering galaxies within $1.5\,r_{500}$.  Among the LoCuSS
sample, neither Abell 1758 or Abell 1914, which are known merging
clusters \citep{okabe}, have high $\fsf$'s, while Abell 611 which
appears to be a regular, relaxed cluster is among the few "active"
clusters with $\fsf{=}0.135$.  Indeed, the most prominent feature in
Fig.~\ref{bo_bcg} is the absence of clusters in the upper right
quadrant of the plot, i.e.\ with large values of $\fsf$ and large
offset between BCG and X-ray peak.  On the face of it, this is counter
to the qualitative expectation that merging clusters contain more
numerous star-forming galaxies than non-merging clusters, as suggested
from the radio observations of merging clusters (not among our sample)
by \citet{miller03}. This may suggest that at least within $r_{500}$
many of the galaxies have already been stripped of their gas when they
were accreted into the progenitor (presumably already massive)
clusters, and so are unable to undergo any starburst phase triggered
by the cluster merger. 

\subsection{Radial population gradients and the infall model}

A comparison of Figs.~\ref{bo_evol}~\&~\ref{bo_r500} indicates that
for many of the clusters, the fraction of mid-infrared sources is
lower within $r_{500}$ than $1.5r_{500}$, in some cases by a factor
two.  This is suggestive of the well known morphology-density and
SF versus density gradients seen in clusters, both at low- and high-redshifts
\citep[e.g.][]{dressler97,balogh,ellingson,treu,smith05a,haines07}.
It has also been noted that the blue galaxy fraction in clusters
depends strongly on the radius within which the measurement is made,
with $f_b$ systematically increasing with cluster-centric radius
\citep[e.g.][]{ellingson,wake}.

\begin{figure}
\plotone{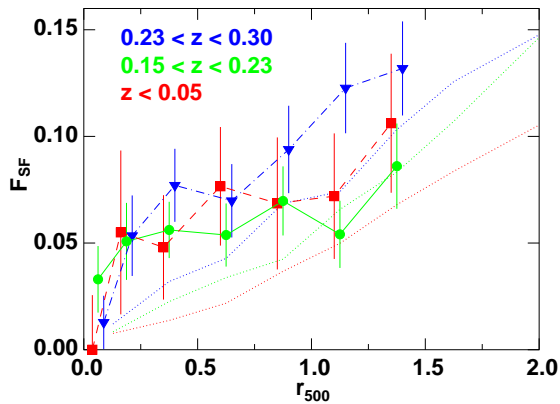}
\caption{Composite radial gradients in the star-forming component for
  clusters with $z{\le}0.05$ (red squares and dashed lines),
  $0.15{<}z{<}0.23$ (green circles and solid lines) and
  $0.23{<}z{<}0.30$ (blue triangles and dot-dashed lines).  For
  comparison the dotted lines indicate the fraction of cluster
  members identified as infalling from the Millennium simulation,
  scaled to match the corresponding field $\fsf$ values at large radii.}
\label{bo_rad}
\end{figure}

In Fig.~\ref{bo_rad} we show the composite radial gradients in the
fraction of mid-infrared luminous galaxies (as defined in
\S\ref{evol}) for clusters in three redshift bins: the seven
$z{\le}0.05$ clusters (red squares and dashed-lines); the low redshift
half of the LoCuSS sample ($0.15{<}z{<}0.23$; green circles and solid
lines); and the high redshift half of the LoCuSS sample
($0.23{<}z{<}0.30$; blue triangles and dot-dashed lines).  The error
bars indicate the uncertainty in the mean value of $\fsf$ for each
sub-sample in each bin and not the cluster-to-cluster scatter.  Each
redshift bin shows a clear increase in $\fsf$ with projected cluster
centric-radius, although within the uncertainties we find no strong
evidence for redshift evolution in the radial trends plotted in
Fig.~\ref{bo_rad}.  However outside $r_{500}$, there appears to be a
modest excess of galaxies in the highest redshift sub-sample.  This
could be due to an increase in the fractional accretion rate of
infalling galaxies with redshift, and/or an increase in the level of
{\em triggered/enhanced} star-formation in galaxies as they encounter
the cluster environment for the first time, such as observed by
\citet{moran05} for galaxies at the virial radius of Cl\,0024+16.
Similarly, \citet{gallazzi} find a significant population of galaxies
undergoing highly-obscured star-formation, preferentially located in
intermediate-density environments typical of those found near the
cluster virial radius, while \citet{fadda08} find enhanced
star-formation activity at $r{>}r_{500}$ along two filaments feeding
Abell~1763.  These trends are also qualitatively consistent with
\citet{ellingson}, who found a steepening in the population gradients
in clusters at $0.30{<}z{<}0.55$ relative to those at $0.18{<}z{<}0.30$,
and can explain the difference in the levels of evolution seen in
Figs.~\ref{bo_evol} and~\ref{bo_r500}.

To test our infall interpretation of the radial trends seen in
Figs.~\ref{bo_evol},~\ref{bo_r500}~\&~\ref{bo_rad} we examined
galaxies falling into 20 massive clusters ($M_{vir}{>}10^{15}\Msol$)
from the Millennium simulation \citep{springel}.  These simulations
cover a $(500h^{-1}\Mpc)^3$ volume, producing DM halo and galaxy
catalogs based on the semi-analytic models ({\sc galform}) of \citet{bower} for
which positions, peculiar velocities, absolute magnitudes and halo
masses are all provided at 63 snapshots to $z{=}0$, allowing the orbit
of each galaxy with respect to the cluster center to be followed.  We
select member galaxies from these twenty clusters that have
$K{<}\Kstar{+}1.5$, and lie within $5\,\Mpc$ of the cluster center, and identify which ones are infalling into the cluster for the first time. 
The fraction of infalling galaxies increases approximately {\em linearly}
with projected radius from close to zero at the cluster center, to
${\sim}2$5\% at $r_{500}$ and ${\sim}5$5\% at $2\,r_{500}$, 
until we reach ${\sim}4\,r_{500}$, 
beyond which we would not expect to find any galaxies that
have passed once through the cluster \citep{mamon}. We find no redshift dependence for $f_{infalling}(r/r_{500})$ within the simulations, at least over the redshift interval $0{<}z{<}0.3$.

We then consider a simple toy model in which the star-formation in these infalling galaxies is instantaneously quenched once they pass through the cluster core. 
 We should not expect that all infalling galaxies are star-forming, a
significant fraction will already be passive either due to
pre-processing within groups or through internal mechanisms such as
AGN feedback. We therefore use the galaxies
from the UKIDSS-DXS fields in \S~\ref{sec:locuss} to estimate the
fraction of galaxies in the field that would satisfy our selection
criteria, obtaining $\fsf{=}0.25{\pm}0.03$ for the $0.15{<}z{<}0.23$ clusters and $\fsf{=}0.29{\pm}0.04$ for the $0.23{<}z{<}0.30$ clusters, and use these values to
re-normalize the radial profile of the Millennium galaxies at
$4\,r_{500}$. For the low-redshift bin, we measure $\fsf$ from the 561 $M_{K}{<}M_{K}^{*}+1.5$ galaxies \citep[taken from the NYU-VAGC;][]{nyuvagc} from the SDSS DR7 having redshifts $z{<}0.1$ also having 24$\mu$m photometry from SWIRE, obtaining $\fsf{=}0.19{\pm}0.02$. 
 Note that we are not using the {\sc galform}-produced SFRs to identify star-forming galaxies, the only model-based parameter we consider is the $K$-band luminosity.

The re-normalized curves are over-plotted on Fig.~\ref{bo_rad} as dotted lines whose color corresponds to its redshift bin. These show the same general trend for $\fsf$ to increase monotonically with cluster-centric radius. 
  This consistency supports our interpretation
of the radial trends in the data as arising primarily from the infall of
star-forming galaxies from the field and suggests that a simple
scenario where infalling star-forming galaxies are quenched once they
pass through the cluster for the first time (for example via
ram-pressure stripping), is valid at least as a first order
approximation to model the evolution of the cluster galaxy population.
The main caveat to this picture is that the observed trends in $\fsf$ appear to lie systematically above the predictions from the simple infall model for all redshift bins over $0.25{<}(r/r_{500}){<}1.5$. One possible cause is projection effects, which can produce dramatic increases in the $f_{SF}$ for {\em individual} clusters, due to the presence of line-of-sight filaments and groups associated with the large-scale structure in which the cluster is embedded, at physical distances 5--20\,Mpc from the cluster, and which are not included in our simple infall model. Alternatively, this excess may indicate star-formation {\em triggered} by environmental processes, or a rather more gradual reduction in star-formation than the instantaneous shut-down in star-formation modeled here. Note that in our model, we make no distinction between those galaxies which pass through the cluster core on almost radial orbits, and those which travel on more circular orbits, never passing within $0.5\,r_{500}$.
In future articles we will use
the spectroscopic redshift information that is becoming available from
our MMT/Hectospec redshift survey to examine in detail the properties
and evolution of the MIR-bright and UV-bright cluster galaxy
populations within the context of the infall scenario.

\section{Summary and conclusions}\label{sec:discuss}

We have presented a study of the mid-infrared properties of galaxies
in a representative sample of 30 massive galaxy clusters over the
redshift range $0{<}z{<}0.4$, taking advantage of panoramic {\em
  Spitzer}/MIPS $24\mic$ and ground-based near-infrared observations
from the LoCuSS and ACCESS surveys.  We revisited the Butcher-Oemler
effect, using our infrared data both to reduce uncertainties on
photometric selection of likely cluster members, and to overcome the
strong dust-obscuration (${\sim}10-30\times$) previously identified in
comparative optical/infrared studies.  We mainly considered
$\fsf$, the fraction of massive cluster galaxies ($K{<}\Kstar{+}1.5$)
within $1.5\,r_{500}$ with $L_{IR}{>}5{\times}10^{10}L_\odot$.  We find that
$\langle\fsf\rangle$ increases steadily with redshift from
$0.035{\pm}0.023$ at $z{<}0.05$, to $0.053{\pm}0.027$ at
$0.15{\le}z{\le}0.25$, and $0.085{\pm}0.038$ at $0.25{\le}z{\le}0.29$,
where the quoted uncertainties are the rms scatter around the means.
This trend is consistent with the trends in $f_{b}$ found
by previous optical studies \citep[e.g.][]{smail,aguerri}, however,
the clusters-to-cluster scatter in $\fsf$ at fixed redshift is roughly
half that seen in $f_b$.  The lower scatter in our results is likely
due to a combination of our use of NIR data to select likely cluster
galaxies and the different physics probed by optical and IR data.

We fit a redshift evolution model of the form $\fsf{\propto}(1+z)^n$ to
the observational data, obtaining a best-fit value of
$n{=}5.7^{+2.1}_{-1.8}$.  This level of evolution exceeds that of
$\Lirstar$ in both clusters and the field \citep[e.g.][]{bai}.  We therefore
repeated our analysis taking into account the cosmic decline in
star-formation by modifying our selection function thus:
$L_{IR}{>}5{\times}10^{10}[(1+z)/(1+z_{\rm max})]^k$, with $k{=}3.2$
chosen to be representative of the results such as those of 
\citet{lefloch} and \citet{zheng}, 
and $z_{\rm max}{=}0.288$, corresponding to the
highest cluster redshift to which we fit the redshift evolution model.
The best-fit redshift evolution for this modified selection is:
$n{=}1.22^{+1.52}_{-1.38}$.  Indeed, if we restrict the galaxy samples
to the central region of each cluster ($r{<}r_{500}$) then the fraction
of cluster galaxies identified as star-forming from their mid-infrared
emission remains constant at ${\sim}5\%$ over $0{<}z{<}0.4$.  This
suggests that redshift evolution of $\fsf$ can be interpreted as a
consequence primarily of the rapid evolution in the SFRs of {\em field} galaxies
over this period, which are accreted onto the clusters at a constant
rate, before being quenched by cluster-related processes.  However the
small residual redshift evolution seen at $r{>}r_{500}$ after removing the
global decline in star formation suggests that some new star-formation
is triggered in clusters at large radii, presumably due to gentle
processes at play in the local group environments within which
galaxies arrive in the clusters.

Globally, there is therefore little evolution in the cluster
population itself, as the infall rate of field galaxies is expected to
remain constant out to $z{\sim}1$ \citep{berrier}, while the
efficiency of the processes which quench star-formation in the
recently accreted galaxies (e.g.\ ram-pressure stripping) should not
evolve rapidly.  This lack of evolution in the cluster galaxy
population to $z{\sim}0.5$ and beyond, is consistent with recent
studies looking at the Butcher-Oemler effect
\citep{smail,depropris03,homeier,wake,andreon} and the morphology-density
relation \citep{holden}. All of these studies indicate that much of the
apparent evolution in earlier optical BO studies was due to the use of
optical luminosities to select galaxies, such that the trends were due
to low-mass galaxies undergoing starbursts in the higher-redshift
samples \citep[see Fig.~2 of][]{holden}, or biases resulting from the
cluster sample itself.

The view that the bulk of star-formation in clusters simply represents
recently accreted field galaxies, is consistent with the observed
constancy of the shape of the UV and IR LFs (both $L^\star$ and
$\alpha$) from clusters to the field in the local Universe
\citep{cortese,bai06}, while \citet{mercurio} find no variation in the
optical LF of blue (i.e.\ star-forming) galaxies with environment.
Moreover, the cluster IR and H$\alpha$ LFs have been
shown to evolve in the same manner as field LFs, declining by a factor
${\sim}10$ since $z{\sim}1$ \citep{bai,finn}.

We have investigated the effects of cluster properties on the level of
star-formation in their member galaxies, finding no apparent
dependence of $\fsf$ on either X-ray luminosity or the dynamical
status of the cluster.  The
absence of correlation with $L_{X}$ may seem to rule out ICM-related
processes as the main route by which star-formation is quenched in
dense environments, but this may simply be a saturation effect, in
that ram-pressure stripping is effective in stripping the gas in all
infalling galaxies even for the lowest X-ray luminosity clusters in
our sample.  Studies which examine much poorer systems find strong
anti-correlations between $\fsf$ and $L_{X}$ or $\sigma_{\nu}$, which
then flatten off for the $\sigma{>}550$\,km\,s$^{-1}$ systems
comparable to those which make up our sample.  The lack of correlation
between $\fsf$ and the dynamical status of the clusters is surprising
and seems to contradict previous studies \citep[e.g.][]{miller03}.
This may suggest that at least within $r_{500}$, galaxies have already
been stripped of their gas, and so are unable to undergo any starburst
phase triggered by the cluster merger.  However at larger radii there may
still be enhanced activity during certain phases of cluster mergers,
as the infalling galaxies should still be gas-rich.  Expanding on
this, composite population gradients show a smooth increase in the
fraction of star-forming galaxies from close to zero in the cluster
cores to 7--13\% by 2\,$r_{500}$.  Through comparison with numerical
simulations, we find these gradual trends are consistent with a simple
model in which the star-forming galaxies are infalling into the
cluster for the first time (usually on highly radial orbits), and then
quenched somehow once they pass through the cluster core, for example
via ram-pressure stripping.  Within $r_{500}$ there is no apparent
evolution in the radial population gradients, but beyond $r_{500}$ we
find a possible excess of $24\mic$-bright galaxies in the highest
redshift bin ($0.23{<}z{<}0.30$), suggestive of either enhanced
star-formation in the cluster infall regions similar to that found by
\citet{moran05}, \citet{gallazzi} or \citet{fadda08}, or an increase in
the fraction of infalling galaxies, comparable to that expected from
simulations \citep{berrier}.

In the future we will further develop these results using spectroscopic
redshifts for the cluster galaxy populations from our ongoing
MMT/Hectospec survey, plus weak-lensing \citep{okabe09} and X-ray
\citep{zhang08} data to investigate in detail the relationship between
the star-forming galaxy populations in clusters and the dynamical
state of the host clusters.

\section*{Acknowledgements}

CPH, GPS, AJRS and RJS acknowledge financial support from STFC.  GPS and
RSE acknowledge support from the Royal Society.  This work was partly
carried out within the FP7-PEOPLE-IRSES-2008 project ACCESS.  We
acknowledge NASA funding for this project under the Spitzer program
GO:40872.  CPH and GPS thank Trevor Ponman and Alastair Edge for
helpful comments on early drafts of this article. We thank our colleagues in the LoCuSS collaboration for their encouragement and help. We also thank the
Virgo Consortium for making the Millennium Simulation available to the
community.

\label{lastpage}
\end{document}